\documentclass[sigconf,anonymous=false]{acmart}
\usepackage{graphicx}
\usepackage{color}
\usepackage{multirow}

\AtBeginDocument{%
  \providecommand\BibTeX{{%
    \normalfont B\kern-0.5em{\scshape i\kern-0.25em b}\kern-0.8em\TeX}}}

\setcopyright{acmcopyright}
\copyrightyear{2021}
\acmYear{2022}
\acmDOI{10.1145/1122445.1122456}

\acmConference[ACM]{}{NY}

\begin{document}

\title{Multi-task Learning with Metadata for Music Mood Classification}

\author{Rajnish Kumar}
\affiliation{%
  \institution{Airtel Digital}
  \country{India}
}

\author{Manjeet Dahiya}
\affiliation{%
  \institution{Airtel Digital}
  \country{India}
}


\begin{abstract}
Mood recognition is an important problem in music informatics and 
has key applications in music discovery and recommendation.
These applications have become even more relevant
with the rise of music streaming.
Our work investigates the research question of whether 
we can leverage audio metadata such as artist and year, 
which is readily available, to improve the performance 
of mood classification models. 
To this end, we propose a multi-task learning approach in which
a shared model is simultaneously trained for 
mood and metadata prediction tasks 
with the goal to learn richer representations.
Experimentally, we demonstrate that applying our technique on 
the existing state of the art convolutional neural networks for
mood classification
improves their performances consistently.
We conduct experiments on multiple datasets 
and report that our approach can lead to improvements
in the average precision metric by up to 8.7 points.
\end{abstract}



\maketitle

\section{Introduction}
\label{sec:introduction}

Music has the ability to express emotions and induce moods \cite{mood-survey}.
The mood of a music piece is said to be the emotion or mood expressed 
by it.
The task of recognizing the mood of a music piece is an 
active research area in music informatics \cite{end-to-end-audio-tagging, essentia,mood-cnn-16,mood-listening-data,mood-survey} and
has multiple applications.

Mood recognition finds key applications in 
the areas of music discovery and recommendation.
Given the rise of music streaming, massive collections of 
music have surfaced online, 
and these research areas have taken a center stage \cite{current-challenges-recsys}.
The capability to detect the moods of a song is
an important requirement for these platforms.
To highlight the importance, 
as a popular music streaming service, 
we get at least 1\% of search queries on our
platform related to
moods such as ``sad songs''.
In other words, users intuitively associate moods with music,
and it is a common user expectation to find music content 
with mood keywords.

Recent work has modeled it as a multi-label classification problem with
deep neural networks,
which has resulted in well-performing models.
The idea is to apply models from vision research 
on audio spectrograms, which could be thought of as visual
representations of audio data 
\cite{mood-cnn-16, essentia, musicnn, musically-motivated1, musically-motivated2}.
More recently, 
researchers have used song representations derived from
listening data as input features instead of spectrograms
for this task \cite{mood-listening-data}.
They have reported interesting insights
that user-song interaction data can be more valuable
than audio data
for predicting the mood of a song.

Surprisingly, previous work has not given much 
attention to the metadata of songs
such as artist and year, 
which is readily available and can potentially be helpful.
In this paper, we propose a method to leverage
this data for improving 
the performance of the existing models for our task.
Specifically, we employ the technique of multi-task learning,
along with audio metadata,
on convolutional neural networks (CNN) for
mood classification \cite{mood-cnn-16, essentia, musicnn}, 
and we find that their performances improve.

Multi-task learning is a training paradigm in which  
multiple tasks are learned simultaneously by a shared model.
The training data of different tasks help in learning internal representations
(i.e., the model layers and parameters) 
that are more informative than the individual models 
of the different tasks
\cite{mtl-vision, mtl-beat-tempo, mtl-instrument-recog, mtl-source-separation}.
In other words, the knowledge present in the data of one task is 
\emph{also} used to improve the other task, and vice-versa.
In our case, we employ multi-task learning approach on two tasks, 
namely
predicting the mood and
predicting the metadata (artist and year),
where the former is our primary task.
The input of the model is the audio spectrograms and 
the output is the labels corresponding
to both the tasks.
We use the same model (i.e., shared parameters) 
and train it simultaneously for both the tasks.

Our work investigates the research question of whether 
we can leverage audio metadata to improve the performance 
of mood classification models. 
To this end, we propose a multi-task learning approach in which
a shared model is simultaneously trained for 
mood and metadata prediction tasks 
with the goal to learn richer representations.
We experimentally demonstrate that applying our technique on 
the existing state of the art CNNs for mood classification 
improves their performances consistently.
We conduct experiments on multiple datasets 
and report that our approach can lead to improvements
in average precision by up to 3.4 points for public
datasets and 8.7 points for 
an internal regional dataset.

The rest of the discussion is organized as follows:
Section~\ref{sec:model} presents the model architecture.
We summarize the datasets and discuss the preprocessing steps in 
Section~\ref{sec:datasets}.
In Section~\ref{sec:exp-eval}, we present the experimental setup
and results.
We discuss related work in Section~\ref{sec:rel-work}, 
and
finally, we conclude in 
Section~\ref{sec:conclusion}.

\section{Model architecture}
\label{sec:model}

Our model is a convolutional neural network (CNN) 
and follows a VGG-like structure \cite{vgg}.
Convolutional neural networks are 
usually applied to analyze visual images.
The idea of applying these
to the audio domain is that 
spectrograms can be thought of 
as
equivalent visual representations of 
audio data containing the relevant information.
The idea is not new, and
significant recent work in music informatics has
borrowed network architectures and insights
from computer vision research
\cite{essentia, musicnn, mood-listening-data, mtl-source-separation, mood-cnn-16, musically-motivated1, musically-motivated2,end-to-end-audio-tagging, mtl-beat-tempo}.

\begin{figure}[htb]
 \centerline{
 \includegraphics[width=0.8\columnwidth]{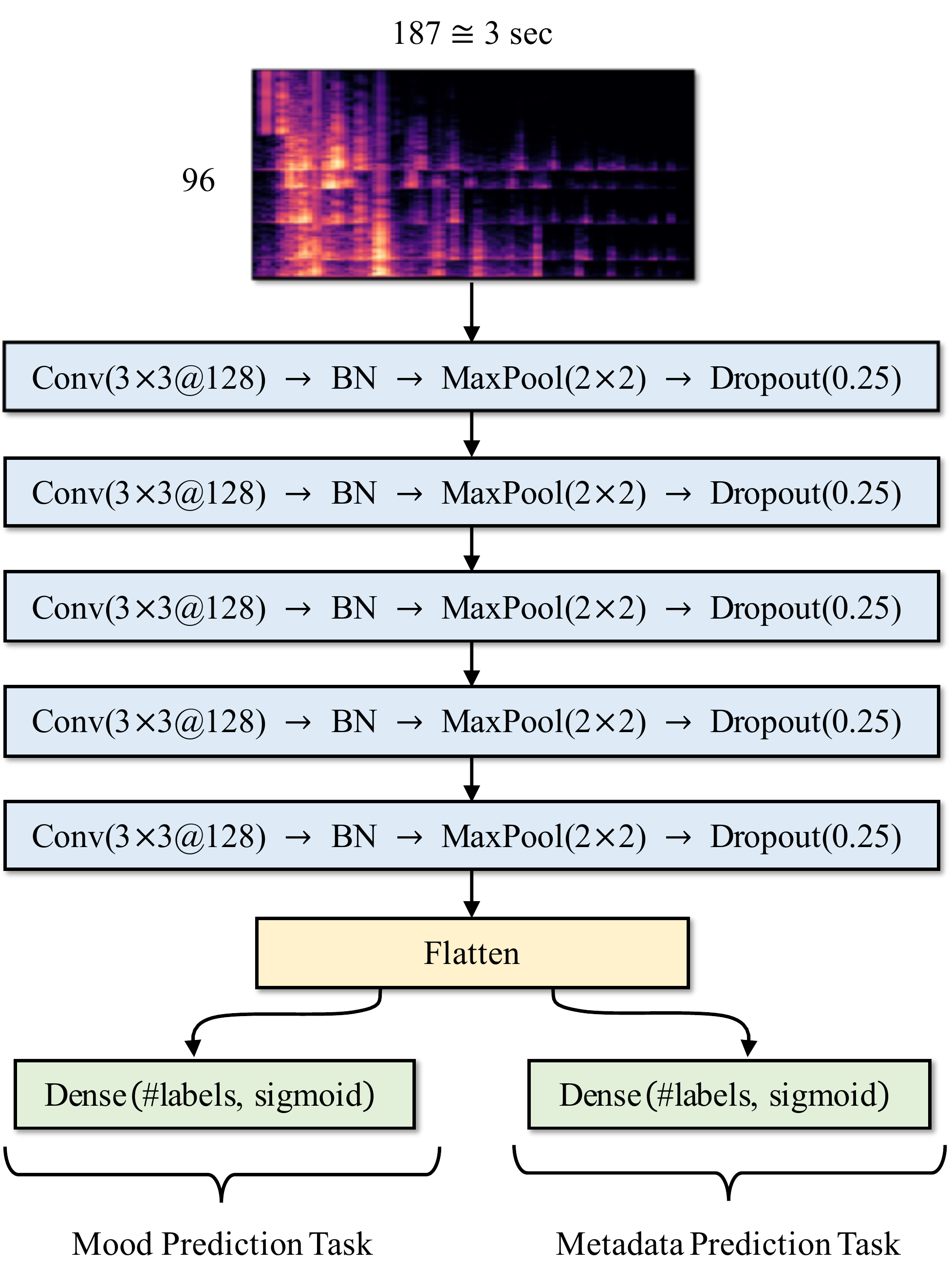}}
 \caption{Multi-task model architecture.}
 \label{fig:model}
\end{figure}

Figure~\ref{fig:model} presents the architecture of our model.
The input of the model is a 2D-tensor of shape $(187 \times 96)$. 
It is supposed to be a mel-spectrogram representation of
an audio sample with 187 frames and 96 mel-bands.
The conversion of audio data to spectrogram is performed 
in a preprocessing step
discussed in Section~\ref{sec:preprocessing}.
As per the configuration of the preprocessing step, 
187 frames of a spectrogram correspond to about 3-sec long 
audio sample.

The input passes through five blocks of layers with exactly the same
configuration.
Each block itself consists of four layers, namely 
convolution,
batch normalization,
max pooling,
and
dropout --- in the same order.
The convolution layer has its filters of size 
(3$\times$3), in other words, 
the filters move in both the axes of input, 
thus, performing a 2D convolution.
The number of filters is 128, the stride is 1,
the padding is set to ``same'', 
i.e., producing convolution output with
the same size as that of the input,
and the non-linearity is set to ReLU.
After the convolution layer, 
batch normalization is applied 
with default parameters \cite{keras-bn}.
Next, max-pooling operation is applied 
with the stride of (2$\times$2) and the pool size of (2$\times$2).
Finally, a dropout layer is applied with 
the dropout rate of 0.25.

After these five blocks, 
a flatten layer is applied to change the shape of the
output tensor from 2D to 1D.
This 1D vector is now passed to be consumed by two different tasks, 
namely mood classification and metadata classification.
Each task is represented as a dense layer, 
with its units equal to the number of labels.
For instance, 
it is the number of different moods for the task of mood classification.
In these dense layers, 
the \textit{sigmoid} activation function is applied to
output valid label probabilities that are mutually-inclusive. 
Thus, it models a multi-label classification task.

The number of trainable parameters in the first five blocks of
the model is 591k.
The number of parameters in the output layers is dependent on the
number of labels in the respective task 
(as well as the number of units in the preceding layer).

The two different output layers
allow us to share the same model
simultaneously for the two different tasks --- 
thus, putting together a setup for \emph{multi-task learning}.
The loss function of the entire task is a weighted sum of 
the losses of the two individual tasks. 
Formally, we train the model with the following loss function:
$$
L = L_{mood} + \alpha L_{metadata}
$$
Here, $L_{mood}$ and $L_{metadata}$ denote the losses corresponding to
mood and metadata prediction tasks, respectively.
The symbol $\alpha$ is a hyperparameter, and
it is the weightage given to the metadata prediction task (secondary task).
Setting $\alpha$ to zero disables multi-task learning, and 
the model degenerates purely into a mood prediction model.
It should be noted that 
metadata is required only at the time of training,
and at the time of prediction, 
we ignore the activations of the metadata output layer.
We discuss the choice of the value of $\alpha$ in
Section~\ref{sec:mtl-weight}.

The model described above works with 
3-sec audio segments and outputs probability values 
for the different mood labels.
A song is mostly longer than that, 
and the inference (at the time of validation and testing) 
for a song is
performed by averaging the results of all 
the 3-sec non-overlapping segments in a song.
For each segment of a song, 
we run the model and compute the probabilities of
the mood labels.
The probability of a specific mood label for a song
is then computed by averaging the probabilities for
this mood label over all the segments of the song.


\section{Datasets}
\label{sec:datasets}

We report our results on three different datasets.
The first two are publicly available, and the third one is 
a private, regional dataset.
Following is a brief summary of the datasets:

\textbf{\textit{MTG-Jamendo dataset (MTG)}} 
contains 18,486 songs with multi-label annotations of
59 moods.
For each song, artist information and
full audio are made available.
The audio data is available in MP3 format with 320 kbps bitrate 
and 44.1 kHz sample rate. 

\textbf{\textit{MagnaTagATune dataset (MTT)}}
contains multi-label tag (including moods) annotations of 
25,877 audio clips of about 30 sec.
The audio clips have been created from
5405 songs provided by the Magnatune label.
For each audio clip, 
over 180 unique tags are available along with artist information.
The audio clips are available in MP3 format with 32 kbps bitrate 
and 16 kHz sample rate.

\textbf{\textit{Internal dataset}}.
Models trained with the above datasets perform 
poorly for songs of regional languages
such as Hindi and Punjabi.
To this end, we manually created our in-house dataset spanning
22,015 songs.
This dataset is annotated with three moods: 
\emph{happy}, 
\emph{sad}, and \emph{energetic}.
We specifically focus on these three tags, as 
these are the most frequently queried mood labels on our platform. 
In addition to mood labels, artist and year information is available.
The audio segments are available in MP3 format with 
up to 320 kbps bitrate
and 44.1 kHz sample rate.

\subsection{Audio preprocessing}
\label{sec:preprocessing}

As a preprocessing step, we convert the audio files to mel-scaled spectrograms.
The generated spectrogram has 96 mel-bands and is rescaled to 16 kHz sample rate.
The frame length is 256, and one second of 
the spectrogram is equal to 62.5 frames.
The converted spectrogram is a 2D matrix with shape $(t \times 96)$, 
$t$ denotes the length of the audio file in the number of frames (proportional to time), 
and 96 is the number of mel-bands.
We use the \emph{librosa} python package \cite{librosa} to implement this 
preprocessing step.
This step is performed once for the entire dataset.
Recall that the model is defined to work with a 2D-tensor containing 
187 frames each with 96 bands, which
is equivalent to a roughly 3-sec long audio file.
Note that for all our experiments, 
we consider only the first 29 seconds of the
audio.


\begin{table*}[]
\centering
\small
\begin{tabular}{c|l | l| l|l|l|l}
\textit{Dataset}                    & \textit{Model type} & $N_{mood}$ & $N_{metadata}$ &\textit{\# songs}                                   & \textit{AUC-ROC (\%)}                                   & \textit{AUC-PR (\%)}                                      \\ \hline \hline
\multirow{6}{*} {\textit{MTG}}      & \textit{Baseline}                                    & 3                                      & -                                   & 4346                                                    & 80.93                                                   & 71.27                                                      \\  
                                   & \textit{Multi-task}                                    & 3                                      & 50                            & 4346                                                    & 83.22 \textbf{(+2.29)}                                           & 74.69 \textbf{(+3.42)}                                              \\ \cline{2-7} 
                                   & \textit{Baseline}                                    & 9                                      & -                                   & 10041                                                   & 49.95                                                   & 13.22                                                      \\ 
                                   & \textit{Multi-task}                                    & 9                                      & 50                            & 10041                                                   & 51.15 \textbf{(+1.20)}                                           & 13.78 \textbf{(+0.56)}                                              \\ \cline{2-7} 
                                   & \textit{Baseline}                                    & 50                                     & -                                   & 17946                                                   & 76.83                                                    & 16.93                                                        \\ 
                                   & \textit{Multi-task}                                    & 50                                     & 50                            & 17946                                                   & 76.81 (-0.02)                                                     & 17.17 \textbf{(+0.24)}                                                        \\ \hline
\multirow{4}{*}{\textit{MTT}}      & \textit{Baseline}                                    & 3                                      & -                                   & 10128                                                   & 95.84                                                   & 93.40                                                      \\  
                                   & \textit{Multi-task}                                    & 3                                      & 50                            & 10128                                                   & 95.99 \textbf{(+0.15)}                                           & 93.95 \textbf{(+0.55)}                                              \\ \cline{2-7} 
                                   & \textit{Baseline}                                    & 50                                     & -                                   & 25863                                                   & 89.89                                                   & 38.21                                                      \\  
                                   & \textit{Multi-task}                                    & 50                                     & 50                            & 25863                                                   & 89.81 (-0.08)                                           & 38.49 \textbf{(+0.28)}                                              \\ \hline
\multirow{3}{*}{\textit{Internal}} & \textit{Baseline}                                    & 3                                      & -                                   & 22015                                                   & 70.45                                                     & 39.78                                                        \\  
                                   & \textit{Multi-task I}                                    & 3                                      & 50                            & 22015                                                   & 74.88 \textbf{(+4.43)}                                        & 47.26 \textbf{(+7.48)}                                                        \\  
                                   & \textit{Multi-task II}                                    & 3                                      & 50+9                  & 22015                                                   & 76.00 \textbf{(+5.55)}                                                     & 48.47 \textbf{(+8.69)}                                                        \\  
                                    \hline
\end{tabular}
\caption{
\footnotesize{
Performance results of \emph{baseline} and \emph{multi-task} models 
over the three different datasets. 
$N_{mood}$ and $N_{metadata}$ represent the number of labels corresponding to
the mood and metadata prediction tasks, respectively.
We perform various experiments by changing these values to create different configurations
within a dataset. 
$N_{metadata}$ = ``-'' denotes that no metadata is used, or otherwise meaning that it is a baseline
without multi-task learning.
Since the number of songs in the training set depends on the number of moods considered,
we also report the \textit{\# songs} for each experiment. 
\textit{AUC} numbers are reported in percentage, 
and \textit{(+x.xx)} denotes the increase in AUC percentage of the respective 
multi-task based model with respect to its baseline.}}
\label{tab:results}
\end{table*}

\section{Experiments and evaluation}
\label{sec:exp-eval}

\subsection{Evaluation metrics}
To measure the performance of a multi-label classification model,
\emph{average precision} (AP) metric is commonly used.
Previous work on mood classification also used the same metric 
\cite{mood-listening-data, essentia, musicnn}.
The average precision of a label (mood) is defined as the weighted average value of
the precision values across different recall values \cite{sklearn-ap-doc}.
In other words, it approximates 
the area of the precision-recall curve, and it is also denoted as
\emph{AUC-PR}.

Since we have multiple labels (i.e., moods), 
we compute the
AP of all the individual moods and then take their average.
This quantity is also called \emph{macro-averaged} AP,
and we report this quantity in our experiments.
It ranges between 0 and 100\% and higher is better.

Additionally, we provide the area of the receiver operating characteristic
curve (ROC), 
as a few papers have reported this metric instead.

\subsection{Training}
We employ batch training with a batchsize of 32 to train our model. 
Since the model takes a spectrogram corresponding to
3-sec long audio as input 
(i.e., 187 frames), 
we require several batches 
with 32 such segments for training.
We build these batches on-the-fly from the spectrograms of the audio files,
which we have computed in the preprocessing step.

The logic for building the batch is as follows:
1) Pick a random song out of all the available songs in the training data.
2) Within the selected song, pick a random window of 187 frames. 
In other words, we pick a random contiguous segment of 3 sec in the selected song.
3) Repeat the steps 1 and 2 to select a total of 32 segments.
4) The target variables,
i.e., the mood and artist labels of a segment is set to 
that of the selected song.

The number of batches per epoch is proportional to the number of songs
in the dataset. Specifically, we generate $1.25 \times num\_songs$ batches
in one epoch, where $num\_songs$ is the number of songs in 
the training dataset.
For all our experiments, we train a model up to 30 epochs, 
with the patience value of 3 epochs for early stopping.

We use the following splits of 
80\%, 10\% and 10\% for training, validation 
and testing datasets, respectively.
We make sure that splits happen at the song level instead of
segments.
It ensures that the segments of the same song are not present 
in the data for training and testing both.
Thus, preventing the problem of data leakage.
For training of our models, 
we use the Adam optimizer \cite{adam} with 
its parameters as follows:
$learning\_rate = 0.001$,  
$beta\_1=0.9$,
$beta\_2=0.999$, and
$epsilon=10^{-7}$.

\subsection{Multi-task learning loss function weight ($\alpha$)}
\label{sec:mtl-weight}
The loss function of the entire task is defined as a linear
combination of the losses of the two individual tasks, that is,
$L = L_{mood} + \alpha L_{metadata}$. 
The hyperparameter $\alpha$ controls the weightage given to the 
metadata prediction task while training.

We experimented with different values of $\alpha$ for 
the MTG dataset to decipher the trend.
We found that the following simple formula for $\alpha$
tends to work in our scenarios:
$$
\alpha = \frac{N_{mood}}{N_{metadata}} 
$$
Here, $N_{mood}$ and $N_{metadata}$ are the number of labels in the
mood and metadata prediction tasks.
In other words, the weight $\alpha$ balances the number
of labels in the two tasks. 
In all our experiments of Table~\ref{tab:results}, 
the value of $\alpha$ is decided by this formula.

\subsection{Results}

We conduct multiple experiments across the three datasets described in 
Section~\ref{sec:datasets}.
For a dataset, we train two types of models: \emph{Baseline} and
\emph{Multi-task}.
Both the model types follow the
architecture of 
Section~\ref{sec:model} with the only difference that 
the latter is trained with the metadata prediction task, and
the former is trained without it.
Comparing the results of
the \emph{Multi-task} model with the \emph{Baseline} model for the same configuration
allows us to see the impact of multi-task learning with metadata
over the baseline.
Moreover, we do this comparative study across the different configurations
of the number of moods and across the different types of metadata considered.
All the results are presented in Table~\ref{tab:results}.

\textbf{\textit{Consistent improvement}}. 
We report that across the datasets and across the configurations, we
consistently get improvements in the AUC-PR metric.
For the \emph{MTG} dataset, we get the improvements of 
3.42, 0.56 and 0.24 points with respect to the baseline models
of respective configurations.
Likewise, for the \emph{MTT} dataset we get 0.55 and 0.28 points improvements.
It should be noted that for the \emph{MTT} dataset, 
the baseline metrics are already too high
and perhaps, because of that 
the gain is small.
Finally, for our \emph{Internal} dataset, we report 
an increase of 8.69 points.
The improvements are also observed in the AUC-ROC metric, except
when the number of mood labels is 50.

Notice that improvements across the datasets are not uniform, 
and they do not follow any correlation.
We believe that the improvements are artifacts of the datasets and
are dependent on the characteristics of the data.

\textbf{\textit{Effect of the number of mood labels}}.
For the \emph{MTG} and \emph{MTT} dataset, 
we also conducted experiments with various values of the 
number of moods considered.
For instance, \emph{MTG} experiment with 3 moods denotes that we
take the top 3 frequent moods in the dataset and train our
model on that, and so on for 9 and 50 moods.
We observe that when the number of mood labels is high, 
the improvements are relatively lower than a configuration with fewer mood 
labels.
For example, the improvements for the \emph{MTG} dataset 
are  
3.42,
0.56
and 0.24 points
for 3, 9 and 50 mood labels, respectively.
Perhaps, the reason for this behavior is that when the number of mood labels
is high, the mood labels data is already discriminative enough that
metadata labels data cannot add much value.
In other words, with a higher number of mood labels, 
the mood data itself is sufficient for learning richer representations.
We also observe this trend for the \emph{MTT} dataset.
However, we do not report these numbers for the \emph{internal} dataset,
as we only have 3 mood labels available in it.

\textbf{\textit{Effect of the types of metadata labels}}.
In all the experiments, we
consider the top 50 frequently occurring artists in the dataset 
to ignore the artists with lower counts.
In case of the internal dataset, we additionally 
consider the ``year'' information of the metadata.
We create 9 classes to represent the years, 
where each class 
is a bucket corresponding to a range of 5 contiguous years.
We train two multi-task models for the internal dataset, 
first (\textit{Multi-task I}) with just 50 artists, 
and the second (\textit{Multi-task II}) 
with 50 artists and 9 classes of years.
The improvements of these models are 7.48 and 8.69 points, respectively.
The ``year'' metadata further improves the metric by 1.21 points.




\section{Related work}
\label{sec:rel-work}

CNN based models have been successfully 
applied to predict mood from the audio data 
\cite{mood-cnn-16, essentia, musicnn}.
These models do not assume any knowledge of the music domain ---
the spectrograms are simply treated as images 
and common vision models are applied on them.
On the other hand, work on musically motivated architectures 
\cite{musically-motivated1, musically-motivated2} 
proposed CNN filters of various shapes to detect
temporal and frequency-based features such as BPM, 
onsets and timbre.
Although these models 
tend to require domain knowledge, 
but the benefit is that
they need fewer parameters 
and perform competitively 
with lesser training data \cite{end-to-end-audio-tagging}.

Filip Korzeniowski et al. explored mood recognition using input features
derived from listening data \cite{mood-listening-data} instead of
spectrogram based audio features.
They reported that features derived from listening data such as
interactions of songs with different users are more
helpful in recognizing moods versus 
not so much available labeled audio data.

It is interesting to note that our work does not make any assumptions
about the model architecture and the input types.
Our simple idea of using audio metadata in a multi-task learning setting
applies to all the aforementioned models and with different kinds of input features.
Moreover, the modifications required by multi-task learning
only demand metadata at the training time, and
it does not disturb inference at all, i.e., it does not
require metadata for inference.
We validate these ideas through
our experiments and 
demonstrate 
that supplementing the existing models with 
multi-task learning (with readily available audio metadata)
improves the performance of a given model.

\section{Conclusion}
\label{sec:conclusion}

We highlighted the importance of the
task of mood recognition from the perspective of a music streaming service.
We reported an observation that 
as a popular music streaming service we get 
more than 1\% of search queries
related to moods.
Thus, making a strong case for us to understand the content 
mood category to be able to provide relevant search results and
to also provide personalized mood-based playlists to the users.

Towards the goal of improving the mood recognition capability,
we investigated a research question of whether 
we can leverage audio metadata such as artist and year, 
which is readily available, to improve the performance 
of mood classification models. 
To this end, 
we proposed a multi-task learning approach in which
a shared model is simultaneously trained for 
mood and metadata prediction tasks 
with the goal to learn richer representations.
We experimentally demonstrated that applying our technique on 
the existing state of the art CNNs for mood classification 
improved their performances consistently.
We conducted experiments on multiple datasets 
and reported that our approach could lead to improvements
in the average precision metric by up to 8.7 points.

\bibliographystyle{ACM-Reference-Format}
\bibliography{main}
\end{document}